\documentclass[aps,showpacs,twocolumn]{revtex4}
\usepackage{mathrsfs}
\usepackage{bm}
\usepackage{amsmath}

\begin{document}
\title{Different mechanisms for efficient optical transmission through bilayered subwavelength patterned metal films}
\author{Jian Wang$^{1}$, Yong Zeng$^{2}$\footnote{The first and second author contributes equally to this article.}, Xiaoshuang Chen$^{1}$\footnote{xschen@mail.sitp.ac.cn}, Wei Lu$^{1}$\footnote{luwei@mail.sitp.ac.cn}, Jerome V. Moloney$^{2}$}
\affiliation{1. National Laboratory for Infrared Physics, Shanghai
Institute of Technical Physics, Chinese Academy of Science, 200083
Shanghai, China \\2. Arizona Center for mathematical Sciences,
University of Arizona, Tucson, Arizona 85721}
\input epsf
\begin{abstract}
Light transmission through bilayered thin metal films perforated
with subwavelength hole arrays are numerically studied based on a
full-vector finite-difference time-domain approach. A variety of
transmission peaks originating from different physical mechanisms
are observed. In addition to the direct tunnelling and
Fabry-P\`{e}rot resonances, generally possessed by idealized
bilayered dielectric slabs, the near-field localized plasmon
polaritons also play important roles. They not only influence the
direct tunnelling in a destructive or constructive way, the
interactions between these localized plasmon polaritons on
different metal films also result in additional channels which
transfer optical energy effectively.
\end{abstract}
\pacs{78.20.Bh} \maketitle

Subwavelength hole arrays in metallic thin films have received
considerable attention recently after the experimental findings of
Ebbesen in 1998 \cite{ebbesen}. The transmission of light through
such a nanostructured system has been shown to be extraordinarily
efficient at specific wavelengths longer than the array period
\cite{bethe,henri,garcia1,garcia2}. Because of many potential
applications such as near-field microscopy and flat-panel
displays, the Ebbesen's work has sparked a wealth of research
activities both experimentally
\cite{matsui,moreno,molen,barnes,bao} and theoretically
\cite{salomon,pendry,webb,ghaemi,lalanne,Popov,muller,shin,yong,Sergey,baida}.
Moreover, various aspects of the single-layer structure have been
studied to exploit the underlying physical mechanisms, including
the arrangement of the subwavelength holes \cite{ebbesen,matsui},
thickness of the metal film \cite{moreno,yong}, polarization of
the incident light \cite{molen}, hole shape \cite{molen} and
symmetry of the whole structure \cite{Sergey}.

A wealth of reasonable explanations about the Ebbesen's
experiments have been presented such as the formation of localized
waveguide resonances \cite{ruan} and shape resonances, as well as
the interference of diffracted evanescent waves
\cite{henri,garcia1,garcia2}. Among them the excitation of
localized plasmon polaritons (LPPs) is widely believed to play an
important role
\cite{salomon,pendry,webb,ghaemi,lalanne,Popov,muller,shin,yong,Sergey,baida}.
LPPs are the oscillations of conduction electrons inside the metal
forced by the external electromagnetic waves. Because of its pure
imaginary wave vector in the normal-to-the-interface direction, it
is therefore a surface wave whose amplitude decays exponentially
with increasing distance into each medium form the interface. This
field confinement leads to the enhancement of the electromagnetic
field at the interface, which in turn leads to the extraordinary
sensitive dependence of the LPPs properties on the surrounding
dielectric conditions \cite{zayats}. In this work, instead of
studying a single-layer structured metal film, we investigate
bilayered films with variable layer separations. We will show that
at least three kinds of physical mechanisms contribute to the
efficient light transmission, including the direct tunnelling,
Fabry-P\`{e}rot resonances, and additional channels formed by the
interaction between the LPPs on different films.

We first consider an idealized structure consisting of two
identical dielectric slabs (with a real refractive index $n_{2}$
and thickness $d$) embedded into a homogeneous medium (with a real
refractive index $n_{1}$). Under normal incidence with wavelength
$\lambda$, the transmission intensity of an individual dielectric
slab is written
\begin{equation}
T_{s}=\left[1+\sin^{2}\phi\left(\frac{n_{1}^{2}-n_{2}^{2}}{2n_{1}n_{2}}\right)^{2}\right]^{-1},
\end{equation}
with $\phi=2\pi n_{2}d/\lambda$. This equation immediately
suggests that (1) light propagates through a homogeneous media
(assuming $n_{1}=n_{2}$) without reflection; (2) the slab
thickness $d$ influences the optical transmission remarkably. More
specifically, $d=n\lambda/2n_{2}$ results in maximal transmission
while $d=(n+1/2)\lambda/2n_{2}$ leads to maximal reflection, with
$n$ being an integer. For the idealized double-layer structure
with $s$ being the layer separation, the corresponding
transmission intensity is
\begin{equation}
T_{d}=\left\{\frac{(2-T_{s})^{2}}{T_{s}^{2}}+\frac{(T_{s}-1)\left[2-2\cos(\alpha+\varphi)\right]}{T_{s}^{2}}\right\}^{-1}
\label{eq2}
\end{equation}
where $\alpha$ stands for the phase of the complex transmission
coefficient of the single slab, and $\varphi=2\pi n_{1}s/\lambda$.
Evidently, $T_{d}=1$ as long as $T_{s}=1$, in other words, an
efficient transmission of single layer also leads to an efficient
transmission of bilayered structure. We denote this mechanism as
direct tunnelling since it only relates to the optical properties
of single-layer slab. Eqs.(\ref{eq2}) further suggests $T_{d}$ is
strongly sensitive to $\varphi$ and therefore the layer separation
$s$, that is, $T_{d}$ oscillates between 1 (with
$\cos(\alpha+\varphi)=-1$) and $T_{s}^{2}/(2-T_{s})^{2}$ (with
$\cos(\alpha+\varphi)=1$). To obtain $\cos(\alpha+\varphi)=-1$,
the separation $s$ must satisfy
\begin{equation}
n_{1}s=\left[n-\frac{1}{2}-\frac{\alpha(\lambda)}{2\pi}\right]\lambda.
\end{equation}
Further assuming the phase $\alpha(\lambda)$ is almost constant in
the wavelength region of interest, the separation $s$ consequently
is proportional to the wavelength $\lambda$. Because this
mechanism is mainly determined by the separation $s$ and similar
to the resonant mechanisms of the Fabry-P\`{e}rot (FP) cavities,
we therefore denote it as FP resonance. It should be mentioned
that the derivative $ds/d\lambda$ of different order resonance
relates to the integer $n$ and therefore has different value.

The freestanding bilayered structured metal films studied consists
of two identical 500-nm-thick silver films, each film is further
perforated by an array of square cross-section holes. The array
period and hole side length are fixed to be 750 nm and 280 nm,
respectively. These two perforated silver films are separated by
vacuum with a variable separation $s$. The relative dielectric
constant of silver is fitted by the Drude model of
\begin{equation}
\epsilon(\omega)=1.0-\frac{\omega_{p}^{2}}{\omega(\omega+i\gamma)},
\end{equation}
where $\omega_{p}=1.374\times 10^{16}$ s$^{-1}$ is the bulk plasma
frequency determined by the density of conduction electrons,
$\gamma=3.21\times 10^{13}$ s$^{-1}$ represents the
phenomenological damping rate \cite{palik}. A similar single-layer
silver film was numerically studied earlier at normal incidence,
and a peak around 802-nm wavelength was observed in the
transmission spectrum \cite{yong,yong1}.

\begin{figure}[t]
\epsfxsize=280pt \epsfbox{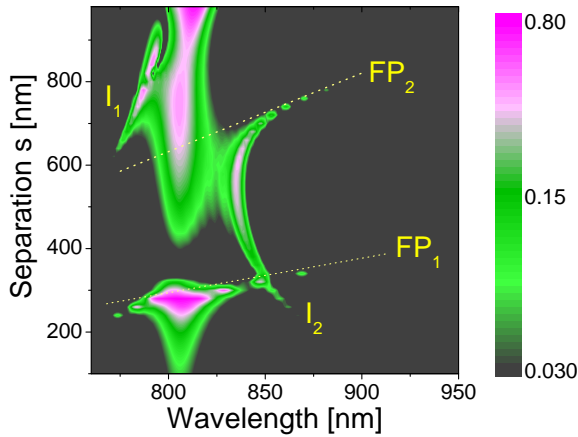} \vspace*{-8.0cm}
\caption{(color online). The transmission spectrum as a function
of the layer separation $s$. The two dashed lines mark the
Fabry-P\`{e}rot-like resonances mainly determined by $s$. I$_{1}$
and I$_{2}$ indicate two additional channels induced by the
interaction between the localized plasmon polaritons on the metal
films.}
\end{figure}

Our main result is shown in Fig.1, where the effect of the layer
separation $s$ on the transmission spectrum of the bilayered
structure under normal incidence is calculated with the aid of a
three-dimensional finite-difference time-domain approach
\cite{taflove}. We are interesting in the wavelength region [750,
950] nm where the far-field diffractions are prohibited at normal
incidence because the wavelength is bigger than the array period.
Multiple transmission peaks occur by increasing the $s$ from 100
nm to 980 nm with an increment of 20 nm, and three important
properties can be observed.

First, the subwavelength patterned metal films can be conceptually
simplified to the idealized bilayered slabs studied earlier by
approximating each perforated film as a homogeneous medium with an
effective permittivity. Consequently physical mechanisms including
direct tunnelling and FP resonances are reproduced to a certain
degree. More specifically, the bright column around 802-nm
wavelength is due to the direct tunnelling, and an example of the
electric-field distribution is shown in Fig.(2a). On the other
hand, the two inclined lines (marked as FP$_{1}$ and FP$_{2}$
respectively) with different slopes are induced by the first-order
and second-order FP resonances, respectively, and their typical
electric-field distributions are plotted in Fig.(2c,2d).
Evidently, as the general characteristics of the FP resonances,
different-order standing waves are formed between the metal films.

Second, the strong overlap or/and repulsion between LPPs around
two metal films lifts up the original degenerate state
(corresponding to $s\rightarrow\infty$ therefore no LPPs
interaction), and results in two additional states for efficient
light transmission \cite{yong1}. Typical electric-field
distributions correspond to the higher-energy (I$_{1}$ in Fig.1)
and lower-energy (I$_{2}$ in Fig.1) states are shown in Fig.(2e)
and Fig.(2f), respectively. Evidently, the higher (lower)-energy
state sits on the left (right)-hand side of the original 802-nm
peak, and the corresponding wavelength should monotonically
increase (decrease) with increasing the film separation $s$
(therefore decreasing the LPPs interaction). Moreover, the LPPs
interaction is so weak for $s$ bigger than 900 nm that these two
states almost coincide with the original 802-nm peak and hence
lead to an enhanced transmission. On the other hand, anti-crossing
between the second-order FP (FP$_{2}$) resonance and the (I$_{2}$)
state happens around a separation of 600 nm. It is possibly
induced by the fact that their modes greatly resemble each other
\cite{sakurai}.

Third, the separation $s$ influences the direct tunnelling
mechanism remarkably in the near-field region ($s<400$ nm). The
transmission intensity first increases steeply from almost zero at
$s=100$ nm to roughly 80\% at $s=280$ nm (the corresponding
electric-field distribution is plotted in Fig.(2b)), then drops to
zero at $s=340$ nm again. Although the FP effects may contribute
somewhat to this rapid variation by tuning the $\textit{cos}$ term
in Eqs.(\ref{eq2}), the dominant mechanism is believed to be the
strong overlap between the near-field LPPs around two metal films.
The LPPs overlap constructively enhancing the transmission in the
region $s<280$ nm, while destructively blocking the light
propagation around $s=340$ nm. Since the overlap is exponentially
decreasing with the increase of the layer separation, its
influence is quite weak for bigger $s$ and the transmission
intensity hence varies smoothly.

\begin{figure}[t]
\epsfxsize=240pt \epsfbox{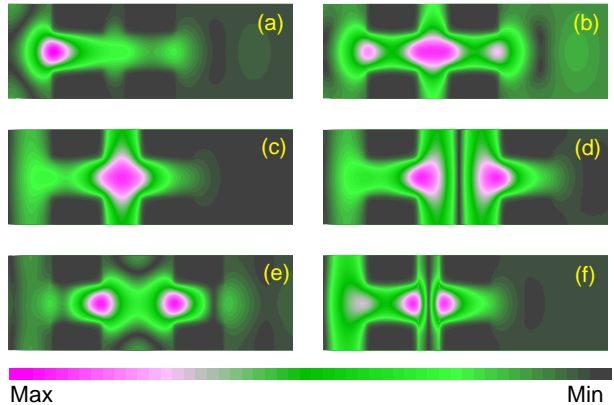} \vspace*{-6.0cm}
\caption{(color online). The distribution of electric field
amplitude corresponds to different mechanisms (see the context),
at cross section (along the light propagation direction) with
different separation $s$. (a) The direct tunnelling with $s=200$
nm; (b) The direct tunnelling with $s=280$ nm; (c) The first-order
Fabry-P\`{e}rot resonance (FP$_{1}$ in Fig.1) with $s=350$ nm; (d)
The second-order Fabry-P\`{e}rot resonance (FP$_{2}$ in Fig.1)
with $s=780$ nm; (e) The higher-energy channel (I$_{1}$ in Fig.1)
with $s=660$ nm; (f) The lower-energy channel (I$_{2}$ in Fig.1)
with $s=200$ nm.}
\end{figure}

To close the discussion, we connect our present result with the
previous studies. It should be emphasized that quite similar
layer-separation-transmission dependence was experimentally
observed in the terahertz (THz) region, and the mechanisms
including both the direct tunnelling and the FP resonances were
found \cite{miyamaru}. Another experiment regarding a cascaded
structured metal films sandwiched a dielectric (vacuum in our
study) layer stressed the contributions of the FP resonance and
the LPPs coupling \cite{ye}. Nanostructured silver multilayers
were also experimentally investigated and the multiple scattering,
the complicated counterpart of the FP resonance, were found to
influence the optical transmission significantly \cite{tang}. Few
studies concentrated on the lateral displacement between the
subwavelength hole arrays in different metal films. Because the
LPPs overlap (interaction) is sensitive to the lateral
displacement, similar to their dependence on the vertical
separation $s$ studied here, the two additional states (I$_{1}$
and I$_{2}$) and even the direct tunnelling can be therefore tuned
by the lateral displacement in the near-field region (small $s$)
\cite{miyamaru,he}.

In summary, we studied the efficient transmission of
electromagnetic wave through subwavelength hole arrays in
bilayered metal thin films by varying the layer separation.
Multiple transmission maximums were observed, and the underlying
physical mechanisms were exploited by comparing the subwavelength
patterned structure with an idealized bilayered dielectric slabs.
The near-field localized plasmon polaritons at the metal films
were found not only to modify the direct tunnelling (in
constructive or destructive way) significantly, but also their
interaction (overlap) form two additional channels for
transmitting the light energy efficiently.

We thank Dr. Colm Dineen for improving the writing. This work is
partially supported by multiple grants from the State Key Program
for Basic Research of China (2007CB613206, 2006CB921704), National
Natural Science Foundation of China (10725418, 10734090 and
60576068), Key Fund of Shanghai Science and Technology
Foundation(08JC14 21100), and Knowledge Innovation Program of the
Chinese Academy of Sciences. Computational resources from the
Shanghai Super-computer Center are acknowledged.

\end{document}